\documentstyle[11pt]{article}
\newcommand{\beq}{\begin{equation}}
\newcommand{\eeq}{\end{equation}}

\newcommand{\beqs}{\begin{eqnarray}}
\newcommand{\eeqs}{\end{eqnarray}}
\newcommand{\beqsn}{\begin{eqnarray*}}
\newcommand{\eeqsn}{\end{eqnarray*}}

\newcommand{\nn}{\nonumber}

\begin{document}

\begin{center}
\vskip 2.5cm
{\LARGE \bf Comment on prescription problem in light-cone gauge}

\vskip 1.0cm
{\Large  D.~K.~Park }
\\
{\large  Department of Physics,  KyungNam University, Masan, 631-701  
Korea}

\vskip 0.4cm
\end{center}

\centerline{\bf Abstract}
 
Recently suggested causal principal value and causal prescriptions
for the "spurious singularity" in light-cone gauge theories are 
nothing but the different guises of usual Mandelstam-Leibbrandt
prescription.
\vfill

\newpage
\setcounter{footnote}{1}

\newcommand{\tr}{\;{\rm tr}\;}
Although the light-cone gauge(radiation gauge in light-cone coordinate) is 
very convenient for the study on QCD vacuum structure and is frequently
used in the various branches[1-4] of physics, from time to time there is a 
debate on the prescription problem for the "spurious singularity" which
appears even in the bare propagator of gauge field. 

It is proved that if the Cauchy principal value(CPV) prescription for 
the "spurious singularity" 
\beq
P_{CPV} (\frac{1}{k^+}) = \frac{1}{2} [\frac{1}{k^+ + i \epsilon} +
                                       \frac{1}{k^+ - i \epsilon} ]
\eeq
is chosen, the calculation of the various Feynman diagrams results in 
double pole which does not make sense physically.[5]
A more reasonable Mandelstam-Leibbrandt(ML) prescription
\beq
P_{ML}(\frac{1}{k^+}) = \frac{k^-}{k^+k^- + i \epsilon}
\eeq
was suggested independently by S.Mandelstam[6] and G.Leibbrandt[7]. Later 
it was proved in the framework of equal-time canonical quantization
that ML prescription preserves the causality[8], and the renormalizibility
of the gauge theories formulated in this way was discussed[9, 10, 11].

Recently Pimentel and Suzuki suggest two kinds of modified prescriptions
for the "spurious singularity", that is causal CPV(CCPV)[12] and causal 
prescriptions[13]. In this short note I will show that these prescriptions
are not new ones but only different forms of ML-prescription.

Their CCPV prescription discussed in Ref.[12] is simply stated as follows:
after usual CPV prescription (1) is chosen, in order to recover the 
causality the constraint that the sign of imaginary part of pole in the
"spurious
singularity" agrees with that in covariant singularity 
$k^2 + i \epsilon = 0$ is required.
This requirement makes change the integration range over the $k^-$ integral
during calculation of various Feynman diagrams and avoid the emergence 
of double pole singularities. They proved these fact by calculating the 
following simple integral
\beq
I = \int \frac{dk}{(k^2 + i \epsilon)[(k-p)^2 + i \epsilon] k^+}.
\eeq
The results of $I$ for the various prescriptions are
\beqs
I_{ML}&=&i (- \pi)^{\frac{d}{2}} \frac{(p^2)^{\frac{d}{2} - 2}}{p^+}
           \Gamma(2 - \frac{d}{2})       \\  \nn
      &\times& \left[ \frac{\Gamma(\frac{d}{2} - 2) \Gamma(\frac{d}{2} - 1)}
                           {\Gamma(d - 3)}
                     - \sum_{l = 0}^{\infty}
                       \frac{(-1)^l \Gamma(2 - \frac{d}{2} + l)}
                            {l! (\frac{d}{2} - 2 + l) \Gamma(2 - \frac{d}{2})}
                       \left(\frac{\vec{p_T}^2}{p^2} \right)^l
               \right],   \\  \nn
I_{CPV}&=& i (- \pi)^{\frac{d}{2}} \frac{(p^2)^{\frac{d}{2} - 2}}{p^+}
           \frac{\Gamma(2 - \frac{d}{2}) \Gamma(\frac{d}{2} - 2) 
                 \Gamma(\frac{d}{2} - 1)}
                {\Gamma(d-3)},        \\  \nn
I_{CCPV}&=& \frac{1}{2} I_{ML},
\eeqs
where $d$ is space-time dimensions and 
$\vec{p_T} = (p_1, p_2, \cdots, p_{d-2})$. The overall factor $\frac{1}{2}$ 
in CCPV prescription is simply explained 
from the fact that their CCPV prescription is nothing but the half
of the usual ML prescription, which can be proved as follows.
The usual CPV prescription with covariant singularity is
\beqs
\frac{1}{k^2 + i \epsilon} P_{CPV}(\frac{1}{k^+})&\equiv&
\frac{1}{2} \frac{1}{k^2 + i \epsilon} 
  \left( \frac{1}{k^+ + i \delta} + \frac{1}{k^+ - i \delta} \right) \\ \nn
                                                 &=&
\frac{1}{4 k^-} \frac{1}{k^+ - \frac{\vec{k_T}^2}{2 k^-} + i \epsilon
                                         \varepsilon(k^-)}
 \left( \frac{1}{k^+ + i \delta} + \frac{1}{k^+ - i \delta} \right)
\eeqs
where $\varepsilon(x) = x / \mid x \mid$.
The requirement that the sign of imaginary part of pole in the "spurious 
singularity" agrees with that in covariant one makes the 
righthand side of Eq.(5) as follows:
\beq
\frac{1}{4 k^-} \frac{1}{k^+ - \frac{\vec{k_T}^2}{2 k^-} + i \epsilon
                                       \varepsilon(k^-)}
  \left[ \frac{\theta(k^-)}{k^+ + i \delta} + 
         \frac{\theta(-k^-)}{k^+ - i \delta}
  \right]
\eeq
where $\theta(x)$ is usual step function.
It is very simple to prove that Eq.(6) is
\beqsn
\frac{1}{2} \frac{1}{k^2 + i \epsilon} P_{ML} ( \frac{1}{k^+} )
\eeqsn
which is the half of usual ML-prescription. Therefore, their CCPV prescription
is nothing but the half of the usual ML prescription.

They suggest another modified prescription in Ref.[13] 
for general non-covariant gauge as follows: for arbitrary vector $n$,
their causal prescription is
\beq
\frac{1}{k \cdot n} \rightarrow \frac{\theta(k^0)}{k \cdot n + i \epsilon} +
                                \frac{\theta(-k^0)}{k \cdot n - i \epsilon}
\eeq
which is reduced to 
\beq
\frac{1}{k^+} \rightarrow \frac{\theta(k^0)}{k^+ + i \epsilon} + 
                          \frac{\theta(-k^0)}{k^+ - i \epsilon}
\eeq
in the light-cone gauge.
Recently they calculated the Wilson-loop by using the prescription (7)[14] 
and it can be shown easily that their result is in agreement with that of
usual ML-prescription in the light cone gauge. However, this is because 
the prescription (8)
is nothing but the ML-prescription. From the fact
\beq
P_{ML}(\frac{1}{k^+}) = P_{CPV}(\frac{1}{k^+}) - i \pi \delta(k^+)
                        \frac{k^-}{\mid k^- \mid},
\eeq
one can easily show
\beq
P_{ML}(\frac{1}{k^+}) = P_{CPV}(\frac{1}{k^+}) - i \pi \delta(k^+)
                        \frac{\alpha k^+ + k^-}{\mid \beta k^+ + k^- \mid}
\eeq
where $\alpha$ and $\beta$ are arbitrary c-numbers because of 
the property of $\delta(k^+)$.
If one chooses $\alpha = \beta =1$, usual ML prescription coincides
with Eq.(8).

Therefore, those prescriptions which were suggested in Ref.[12, 13] are not
new ones but only the different forms of usual ML-prescription.

\newpage
\noindent {\em ACKNOWLEDGMENTS}

This work was carried out with support from the Korean Science and 
Engineering Foundation, 961-0201-005-1.


\begin{thebibliography}{9}

\bibitem{1} I.Brink, O.Lindgren, and B.E.W.Nilsson, Nucl. Phys. {\bf B212}, 401(1983).
\bibitem{2} M.A.Namazie, A.Salam, and J.Strathdee, Phys. Rev. {\bf D28}, 1481(1983).  
\bibitem{3} M.B.Green, and J.H.Schwarz, Nucl. Phys. {\bf B181}, 502(1981).
\bibitem{4} M.B.Green, and J.H.Schwarz, Nucl. Phys. {\bf B198}, 252(1982). 
\bibitem{5} D.M.Capper, J.J.Dulwich and M.J.Litvak, Nucl. Phys. {\bf B241}, 463(1984). 
\bibitem{6} S.Mandelstam, Nucl. Phys. {\bf B213}, 149(1983). 
\bibitem{7} G.Leibbrandt, Phys. Rev. {\bf D29}, 1699 (1984). 
\bibitem{8} A.Bassetto, M.Dalbosco, I.Lazzizzera, and R.Soldati, Phys. Rev.
{\bf D31}, 2012(1985).  
\bibitem{9} A.Bassetto, M.Dalbosco, and R.Soldati, Phys. Rev. {\bf D36}, 
3138(1987).
\bibitem{10} C.Acerbi and A.Bassetto, Phys. Rev. {\bf D49}, 1067(1994).
\bibitem{11} A.Bassetto, hep-th/9605421.
\bibitem{12} B.M.Pimentel and A.T.Suzuki, Phys. Rev. {\bf D42}, 2115(1990).
\bibitem{13} B.M.Pimentel and A.T.Suzuki, Mod. Phys. Lett. {\bf A6}, 2649(1991).
\bibitem{14} B.M.Pimentel and J.L.Tomazelli, hep-th/9602168.
 
\end{thebibliography}
\end{document}